\documentclass[useAMS,usenatbib]{mn2e}

\usepackage{graphicx}
\usepackage{amssymb}

\usepackage{natbib,times}
\usepackage{lscape}

\newcommand{\mc}{\multicolumn}

\begin{document}

\title[Luminosity function of cluster galaxies]
      {Dependence of the bright end of galaxy luminosity function on cluster dynamical state}

\author[Z. L. Wen and J. L. Han]
{Z. L. Wen\thanks{E-mail: zhonglue@nao.cas.cn} 
and J. L. Han 
\\
National Astronomical Observatories, Chinese Academy of Sciences, 20A
Datun Road, Chaoyang District, Beijing 100012, China\\
}

\date{Accepted 2014 ... Received 2014 ...}

\pagerange{\pageref{firstpage}--\pageref{lastpage}} \pubyear{2014}

\maketitle

\label{firstpage}


\begin{abstract}
Luminosity function of cluster galaxies provides a fundamental
constraint on galaxy evolution in cluster environments. By using the
bright member galaxies of a large sample of rich clusters identified
from Sloan Digital Sky Survey, we obtain the bright end of
composite luminosity functions of cluster galaxies, and study their
dependence on cluster dynamical state.
After a redshift-evolution correction of absolute magnitude, the
luminosity function of member galaxies can be well fitted by a
Schechter function when the brightest cluster galaxies (BCGs) are
excluded. The absolute magnitudes of BCGs follow a Gaussian
function with a characteristic width of about 0.36~mag. We find
that the luminosity function of galaxies in more relaxed clusters has
a fainter characteristic absolute magnitude ($M_{\ast}$), and these
clusters have fewer bright non-BCG member galaxies but a brighter BCG.
Our results suggest the co-evolution of galaxy population with cluster
dynamical state and somewhat support the hierarchical formation
scenario of the BCGs.
\end{abstract}


\section{Introduction}

Clusters of galaxies are the most massive bound systems in the
universe, which were formed hierarchically by accretion and merger of
smaller sub-clusters and groups \citep[e.g.][]{cwj+99}. They are
important laboratories to investigate the formation and evolution of
galaxies in dense environment \citep{bo84,gyf+03,glp+04,db07}. The
population of cluster galaxies in the local universe is dominated by
red sequence galaxies. The hierarchical and the passive evolution
models are two important scenarios on the evolution of cluster
galaxies. The hierarchical model \citep[e.g.][]{dsw+06} predicts that
more massive cluster galaxies have a history of earlier star formation
and later stellar mass assembly. About the half of stellar mass in the
most massive galaxies is assembled at a redshift of $z<0.8$ through
merger process. In contrast, the passive evolution model implies that
cluster galaxies were formed in a rapid starburst at very early time
of the universe, and evolved later without any star formation and
merger \citep{dse+99,dse+07}. Luminosity function of cluster galaxies
provides a fundamental constraint on galaxy evolution in cluster
environments \citep[e.g.][]{lmg+06,cbh09,rlr13}.

In general, the luminosity function of galaxies in clusters is defined
as being the number density of galaxies per absolute magnitude as a
function of luminosity, which can be fitted by a Schechter function
\citep{sch76}:
\begin{eqnarray}
\phi_s(M)dM&=&0.4\ln(10)\phi_{\ast}10^{-0.4(M-M_{\ast})({\alpha+1})}\nonumber\\ 
&\times&\exp\Big[-10^{-0.4(M-M_{\ast})}\Big]dM,
\end{eqnarray}
where $\alpha$ is the faint-end slope, $M_{\ast}$ is the
characteristic absolute magnitude, and $\phi_{\ast}$ is the
normalization factor. The luminosity of the brightest cluster galaxy
(BCG) in each cluster is very different from other cluster
galaxies. The luminosity distribution of a sample of BCGs follows a
Gaussian function \citep{hmw+05,hsw+09,dpl+11}.

To understand galaxy evolution in cluster environments, many efforts
have been made to search for the changes of galaxy luminosity
functions with the properties of whole clusters (e.g. redshift,
cluster mass and dynamical state) or member galaxies. Clusters with a
cD galaxy have a significantly different galaxy luminosity function
from spiral-rich clusters \citep{oem74}. The luminosity function of
galaxies in rich clusters has a brighter $M_{\ast}$ and a steeper
$\alpha$ than that in poor clusters
\citep[e.g. ][]{lms04,hmw+05}. Galaxy population and the luminosity
function vary with distance to the cluster center
\citep[e.g.][]{hmw+05}, and the value of $\alpha$ is steeper in the
outer region than that in the central region \citep{dpl+11}. The
luminosity function of early-type galaxies has a flatter $\alpha$ than
that of late-type galaxies \citep{gom+02,myh+07}. The $\alpha$ value
for red cluster galaxies at high redshifts may be
\citep[e.g.][]{dpa+04,tmr+04,sse+07,rvp+09}, or may be not
\citep[e.g.][]{cbh09,mbg+12,dpb13}, smaller than that for local
cluster galaxies. Simple pure passive evolution was claimed by
comparing luminosity functions of galaxies in clusters up to redshift
$z\sim 1$ \citep{dse+99,dse+07,lmg+06,cbh09}, which is inconsistent
with the hierarchical model.

Many clusters have experienced recent merger and show an unrelaxed
dynamical state \citep[e.g.][]{bpa+10,wh13}. Relaxed clusters may have
fewer bright member galaxies than unrelaxed clusters
\citep{dre78,bgb+12}.  However, there is no consensus on the possible
relation between galaxy luminosity distributions and cluster dynamical
states. Galaxy luminosity functions of some individual merging
clusters can not be described by a single Schechter function but by a
double Schechter function \citep[e.g.  A209 and
  A168,][]{mmm+03,yzy+04} or by the superposition of a Schechter
function and a Gaussian function \citep[e.g. the Coma cluster,
  by][]{bdg+95}. \citet{byl07} found a weak correlation between
$M_{\ast}$ and the cluster Bautz-Morgan classification, and the later
is related to cluster dynamical state \citep{wh13}. The luminosity
function of galaxies in clusters with a Gaussian velocity distribution
(i.e. in a relaxed state) has a brighter $M_{\ast}$ and a steeper
$\alpha$ than that in the non-Gaussian clusters \citep{rlr13}, which
suggests again that the luminosity function of cluster galaxies is
really related to cluster dynamical state. However, no significant
difference was found between luminosity functions of galaxies in
clusters with different Bautz-Morgan classifications \citep{col89} or
in clusters with and without substructures \citep{dcd+03,dpb13}. The
discrepancy of these results may come from the limited number of
galaxies of a small number of clusters in previous investigations.

To check if there is any dependence of galaxy luminosity function on
cluster dynamical state, the member galaxy data of a large sample of
clusters with quantified dynamical states are needed. Previously
qualitative classifications for relaxed or unrelaxed (or X-ray
cool/non-cool) clusters \citep{bfs+05,vmm+05,crb+07} are too crude for
such a study. Only a few clusters have their dynamical state carefully
quantified by substructures in X-ray images
\citep[e.g.][]{bt95,bpa+10,wbs+13}.  Recently, we used the photometric
data of Sloan Digital Sky Survey (SDSS) to quantify the dynamical
states for 2092 rich clusters \citep{wh13}, which is currently the
largest cluster sample available with quantified dynamical state. In
this paper, we use this cluster sample to investigate the dependence
of the bright end of galaxy luminosity function on the cluster
dynamical state. In Section 2, we introduce the cluster sample, the
member galaxy data for composite luminosity function, and the
quantified parameter for cluster dynamical state. In Section 3, we
show the dependence of the bright end of galaxy luminosity
function on cluster dynamical state. We present discussion and
conclusion in Section 4.

Throughout this paper, we assume a $\Lambda$CDM cosmology, taking
$H_0=$100 $h$ ${\rm km~s}^{-1}$ ${\rm Mpc}^{-1}$, with $\Omega_m=0.3$
and $\Omega_{\Lambda}=0.7$.

\section{Cluster sample and luminosity function of bright member galaxies}

Using photometric redshifts of galaxies, we identified 132,684 galaxy
clusters from the SDSS DR8 \citep{whl12}, which is an update of the
previous catalog as made from SDSS DR6 \citep{whl09}. Clusters were
identified if they have a richness of $R_{L\ast} \ge 12$ and the
number of member galaxies $N_{200} \ge 8$ within a radius of
$r_{200}$. Here, $r_{200}$ is the radius within which the mean density
of a cluster is 200 times of the critical density of the universe. The
cluster richness is defined as $R_{L\ast}=L_{200}/L^{\ast}$, i.e., the
$r$-band total luminosity of member galaxies of $M^e_r\le-20.5$ within
$r_{200}$ in units of $L^{\ast}$, where $M^e_r$ is $r$-band absolute
magnitude after passive evolution being corrected (see below). 

With these clusters, we need to know their dynamical state for this
work. Three-dimensional distribution and motions of the member
galaxies or hot intracluster gas are the most direct tracer of
dynamical state of clusters, which show several observable effects,
either the velocity distributions in the radial direction or the
galaxy distribution or the gas distribution on the projected sky
plane. The relaxed clusters of galaxies should show a Gaussian
distribution of the radial velocities, and the unrelaxed clusters show
non-Gaussian velocity peak in optical spectroscopic data for member
galaxies \citep{cd96,hmp+04}. However, spectroscopic observations
usually are incomplete for cluster member galaxies and only available
for a very limited sample of galaxy clusters. On the other hand, the
unrelaxed or merger clusters usually show asymmetric distribution of
member galaxies or hot gas. The dynamical state of galaxy clusters can
be derived from the gas distribution by using substructures in X-ray
images for small samples of galaxy clusters, e.g. quantitatively by
using the power ratio \citep[e.g.][]{bt95,bpa+10}, the centroid shift
\citep[e.g.][]{mef+95,mjf+08}, the asymmetry and the concentration
\citep[e.g.][]{hbh+07,srt+08}. Currently, only a few hundred nearby
clusters have their substructures quantified from X-ray image or
optical spectrometry \citep[e.g.][]{ds88,bt95,wbs+13}.

Recently, we presented a method to diagnose the substructure and
quantify the dynamical state of rich galaxy clusters by using
photometric data of the SDSS \citep{wh13}. For each cluster, member
galaxies were selected to have an evolution-corrected magnitude of
$M^e_r \le -20.5$~mag. We constructed an optical smoothed map by
convolving the brightness distribution of member galaxies with a
Gaussian kernel. The asymmetry factor $\alpha$, the ridge flatness
$\beta$, and the normalized deviation $\delta$, were then calculated
from the smoothed optical map. Based on these three parameters, a
relaxation parameter $\Gamma$ was defined to quantify dynamical state
of clusters, which have been optimized by using a sample of 98
clusters with qualitatively known dynamical states of `relaxed' and
`unrelaxed' in literature. A larger value of $\Gamma$ indicates the
more relaxed state of a cluster. The defined $\Gamma$ can successfully
separate 94\% known `relaxed' and `unrelaxed' clusters, and has very
tight correlations with substructure parameters obtained from X-ray
data \citep[e.g.][]{bfs+05,ceg+10}. With these tests and comparisons,
we believe that our 'relaxation parameter' deduced in \citet{wh13}
from photometric data can reliably quantify cluster dynamical
state. Applying this method, we calculated the relaxation parameter
$\Gamma$ for 2092 clusters from \citet{whl12} with a richness
$R_{L\ast}\geq50$ in the redshift range of $0.05<z< 0.42$. The
redshift range is selected to make the cluster sample and also bright
member galaxies to be approximately volume-limited complete
\citep{whl12}. Above the richness of $R_{L\ast}=50$, clusters have
enough bright member galaxies to get a reliable relaxation parameter
$\Gamma$. The values of $\Gamma$ are continuously distributed in the
range of $-2\lesssim\Gamma<0.6$ \citep{wh13}. The sample of 2092
clusters is the largest available with quantified dynamical state, and
therefore is used in this paper to calculate the bright end of galaxy
luminosity function.

\begin{figure}
\centering
\includegraphics[width = 70mm]{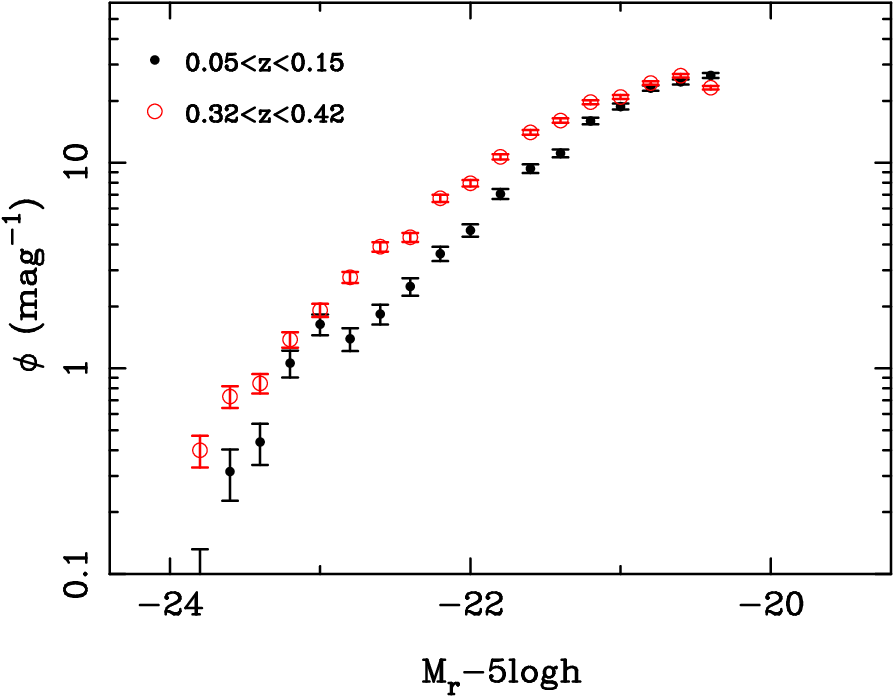} \\[1mm]
\includegraphics[width = 70mm]{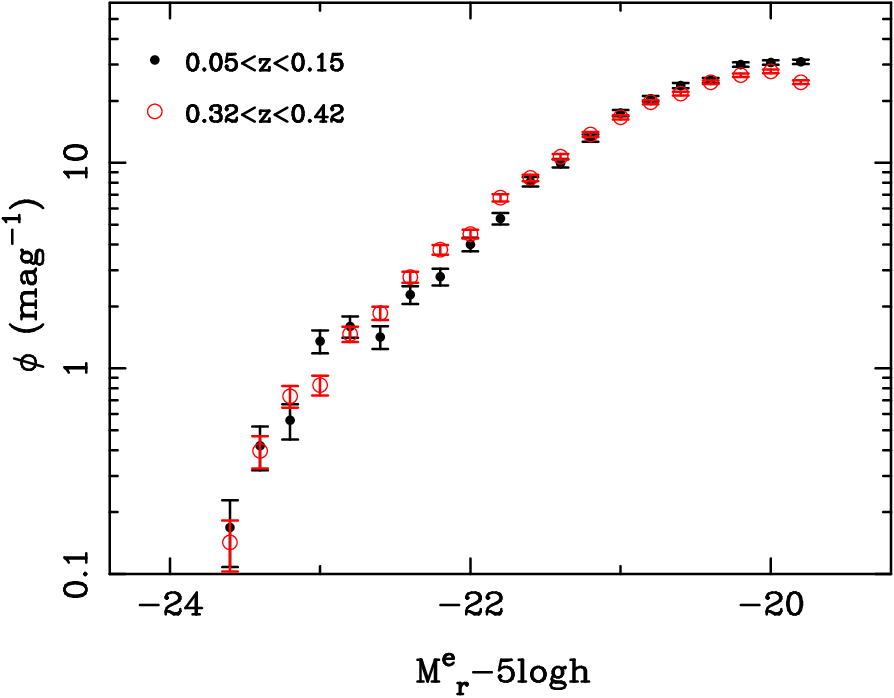}
\caption{Luminosity function of member galaxies within $r_{200}$ of
  rich clusters in two redshift ranges before (upper) and after
  (lower) evolution correction of the absolute magnitude for member
  galaxies via $M^e_r=M_r+1.16\,z$.}
\label{evo}
\end{figure}

We recognize the member galaxies of the 2092 rich clusters by using
photometric redshifts from the SDSS DR8. Because the star/galaxy
separation is reliable to $r=21.5$~mag for the SDSS photometric data
\citep{lgi+01}, the member galaxies are complete down to the limit
$M_r=-20.3+5\log h$ within $z<0.42$ \citep[see][]{whl12}. For each
cluster, the member galaxies are extracted if they have a photometric
redshift within $0.04(1+z)$ from the cluster redshift. For such
bright galaxies, this photometric redshift range was chosen to
include $\sim90\%$ member galaxies but with only $\sim$10\%--15\%
contamination for rich clusters \citep[see][]{whl09}. To further
diminish the contamination of member galaxies and reduce the member
incompleteness, we complement the photometric data with the
spectroscopic redshifts of the SDSS DR10 \citep{dr10+13} for member
galaxies. The galaxies are discarded from the member galaxy list if
they have a velocity difference of $\Delta v>2500$ km~s$^{-1}$ in the
rest frame from the spectroscopic redshift of a cluster. We also
include the missing member galaxies into the photometric redshift data
if their spectroscopic redshifts are within a velocity difference of
$\Delta v\le2500$ km~s$^{-1}$. The galaxies within $r_{200}$ are
considered as member galaxy candidates of the cluster. For background
subtraction, the galaxies between 2 and 4 Mpc from the cluster center
and fainter than the second BCG are considered as being background
galaxies, because the recognized BCG is always considered as member
galaxy of a cluster.

We use these bright member galaxies to derive the bright end of a
composite luminosity function following the method of
\citet{col89}. The number of galaxies in the $j$th bin of the
composite luminosity function is
\begin{equation}
N_{cj}=\frac{N_{c0}}{m_j}\sum_i\frac{N_{ij}}{N_{i0}},
\end{equation}
where $N_{ij}$ is the number in the $j$th bin of the $i$th cluster
luminosity function after background subtraction, $N_{i0}$ is the
normalization of the $i$th cluster, and
\begin{equation}
N_{c0}=\sum_i N_{i0},
\end{equation}
$m_j$ is the number of clusters contributing to the $j$th bin. We only
consider the bright end of galaxy luminosity function in the absolute
magnitude range where the member galaxies are approximately
volume-limited complete, so that $m_j$ is the total number of
clusters. The error of the number in the $j$th bin is
\begin{equation}
\delta N_{cj}=\frac{N_{c0}}{m_j}\Big[\sum_i\Big(\frac{\delta N_{ij}}{N_{i0}}\Big)^2\Big]^{1/2},
\end{equation}
where $\delta N_{ij}$ is determined by the Poisson statistics. The
faint galaxies with a lower luminosity are not considered here because
many of them are late-type (spiral or irregular) and have a larger
uncertainty on the estimated photometric redshift. The recognization
of faint member galaxies is not as complete as bright galaxies, which
may induce bias at the faint end of luminosity function. As pointed
out by \citet{doe+03} and \citet{pdd+04}, the clustering of background
galaxies may induce uncertainty on galaxy number count. Nevertheless,
clustering uncertainty is much smaller than the Poisson error at the
bright-end though hence can be ignored.

The normalization of the composite luminosity function by the method
of \citet{col89} depends on the total number of clusters. It is not
obvious to show in a figure the difference of the composite luminosity
functions between the subsamples of clusters with different redshifts
or dynamical sates. In this paper, we define a normalized composite
luminosity function by dividing the $N_{cj}$ (and similarly for
$\delta N_{cj}$) by the total number of clusters together with the
width of absolute magnitude bin ($\Delta M_r$)
\begin{equation}
\phi_j=\frac{N_{cj}}{m_j\Delta M_r}.
\end{equation}

Some of previous studies showed that the evolution of member galaxy
population can be described by a passive evolution model over a wide
range of redshift \citep{lmg+06,dse+07,cbh09}, which means that galaxy
population becomes older and fainter at lower redshifts. When taking
member galaxies of a number of clusters over a wide range of redshift
for a composite luminosity function, the evolution effect must be
eliminated. As shown in Figure~\ref{evo}, member galaxies within
$r_{200}$ of clusters at higher redshifts ($0.32<z<0.42$) are
systematically brighter than those at lower redshifts
($0.05<z<0.15$). Here we take a linear form of the redshift evolution,
and define an evolution-corrected magnitude,
\begin{equation}
M^e_r=M_r+Q\;z,
\end{equation}
where $Q$ is the evolution slope. Assuming that the member galaxies
were formed in a single burst at the epoch of about $z_f=2$
\citep{lmg+06,cbh09}, we apply a stellar population synthesis model
\citep{bc03} with the initial mass function of \citet{cha03} and solar
metallicity, and we find the value of $Q=1.16$. After the
redshift-evolution correction of the absolute magnitude, the
luminosity functions in different redshift ranges become roughly
consistent (see lower panel of Figure~\ref{evo}). In the following
analysis, we use $M^e_r$ to calculate the composite galaxy luminosity
function to the absolute magnitude limit of $M^e_r=-19.7+5\log h$ over
a wide redshift range of $0.05<z<0.42$.

\section{Dependence of the bright end of galaxy luminosity function on cluster dynamical state}

We use the sample of 2092 rich clusters of $R_{L*}>50$ with known
dynamical states quantified by \citet{wh13} to exam the dependence of
galaxy luminosity function on cluster dynamical state. Here we
  emphasize that we only work at the bright end. Because the
luminosity of a BCG is very distinct from non-BCG member galaxies
\citep[e.g.][]{hmw+05}, we study their composite luminosity
functions separately.

\begin{figure}
\centering
\includegraphics[width = 72mm]{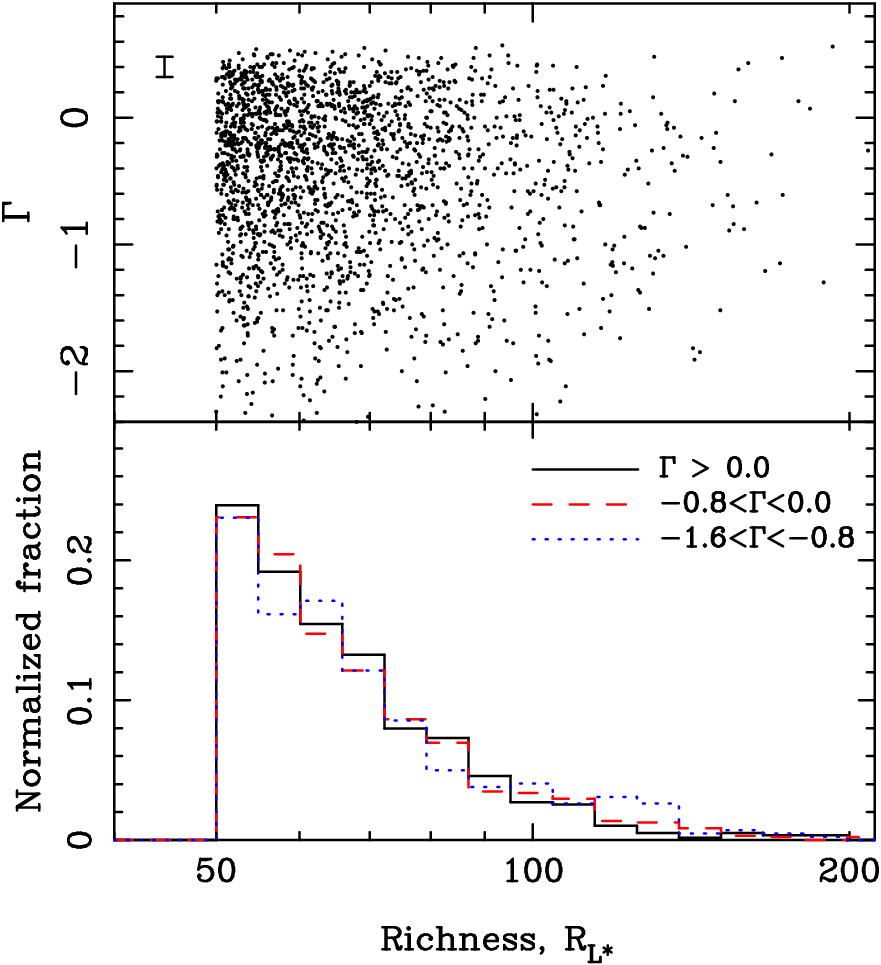}
\caption{The richness distributions are almost the same for the
three subsamples with different ranges of relaxation parameters.}
\label{relax_rich}
\end{figure}

\begin{table*}
{\footnotesize
\begin{center}
\caption{Best-fit parameters of luminosity functions of member galaxies
in clusters with three ranges of relaxation parameters}
\label{tab1}
\begin{tabular}{ccccccc}
\hline
\mc{1}{c}{Relaxation parameter} & No. of clusters &\mc{1}{c}{$\phi_{\ast}$} &\mc{1}{c}{$M_{\ast}-5\log h$} &
\mc{1}{c}{$\phi_0$} & \mc{1}{c}{$M_0-5\log h$} & \mc{1}{c}{$\sigma_0$}  \\
\mc{1}{c}{(1)} &\mc{1}{c}{(2)} &\mc{1}{c}{(3)} & \mc{1}{c}{(4)} & \mc{1}{c}{(5)} & \mc{1}{c}{(6)} & \mc{1}{c}{(7)}  \\
\hline
$ \Gamma > 0.0$    & 589 & 36.8$\pm0.9$& $-20.93\pm0.02$& 2.8$\pm0.3$& $-23.08\pm0.04$& 0.39$\pm0.02$\\
$-0.8< \Gamma< 0.0$& 949 & 32.5$\pm0.5$& $-21.13\pm0.02$& 3.0$\pm0.2$& $-22.66\pm0.02$& 0.36$\pm0.02$\\
$-1.6< \Gamma<-0.8$& 421 & 32.7$\pm0.8$& $-21.20\pm0.02$& 3.2$\pm0.3$& $-22.43\pm0.02$& 0.35$\pm0.01$\\
\hline
\end{tabular}
\end{center}
Notes: Column (3) and (4) are the best-fit parameters of the Schechter
function with a fixed faint-end slope of $\alpha=-1.0$ for the non-BCG
member galaxies; Column (5)--(7) are the best-fit parameters of the
Gaussian function for the BCGs.}
\end{table*}

\begin{figure}
\centering
\includegraphics[width = 72mm]{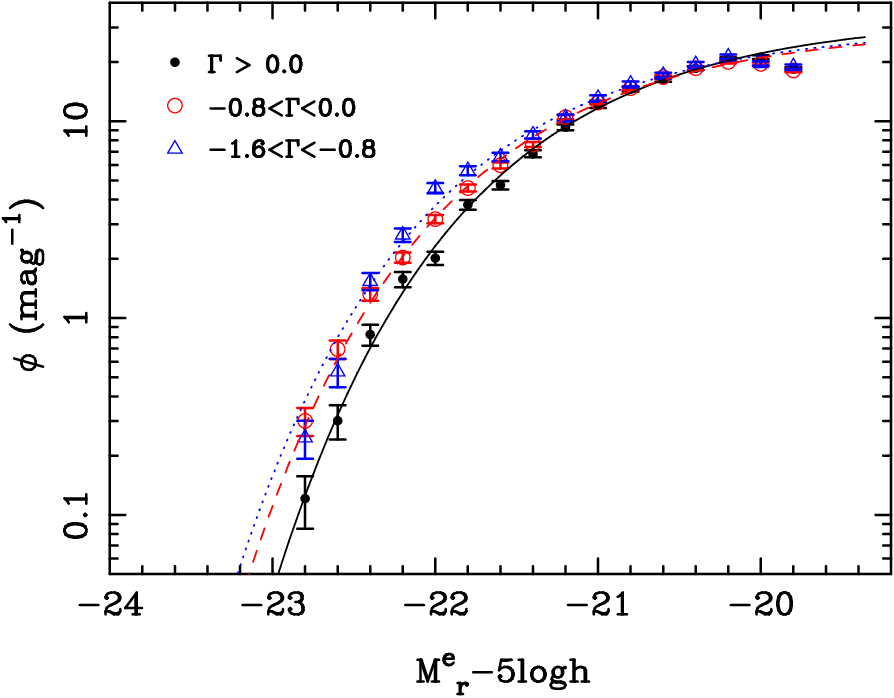}
\includegraphics[width = 72mm]{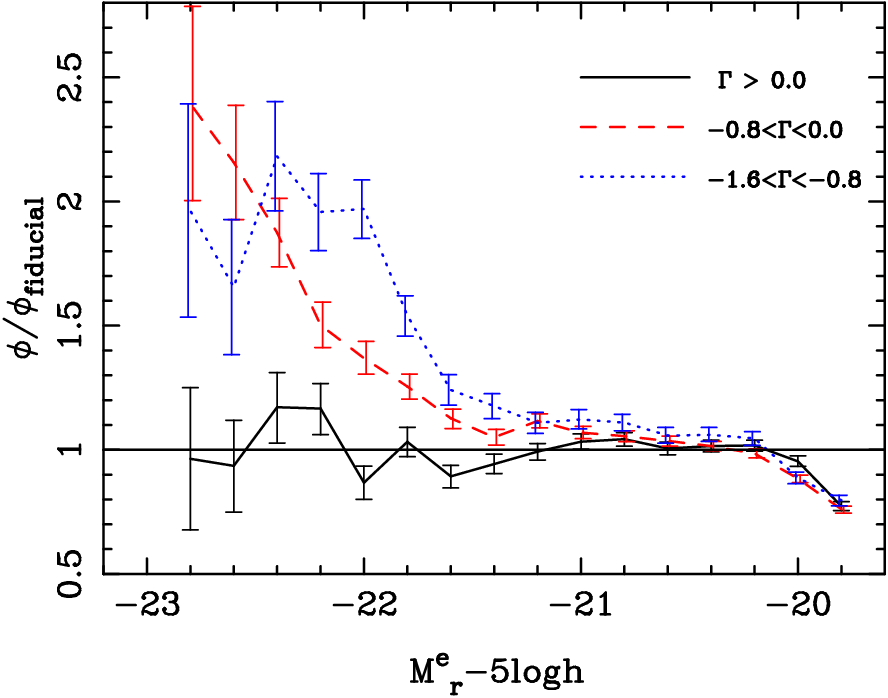}
\caption{Composite luminosity functions and the best-fit Schechter functions
  ({\it Upper panel}) and the ratio to the fiducial line ({\it Lower
    panel}) for member galaxies within $r_{500}$ of clusters in three
  ranges of $\Gamma$. The fiducial line is the best-fit Schechter
  function of member galaxies in relaxed clusters of $\Gamma>0.0$.}
\label{relax}
\end{figure}

\begin{figure}
\centering
\includegraphics[width = 72mm]{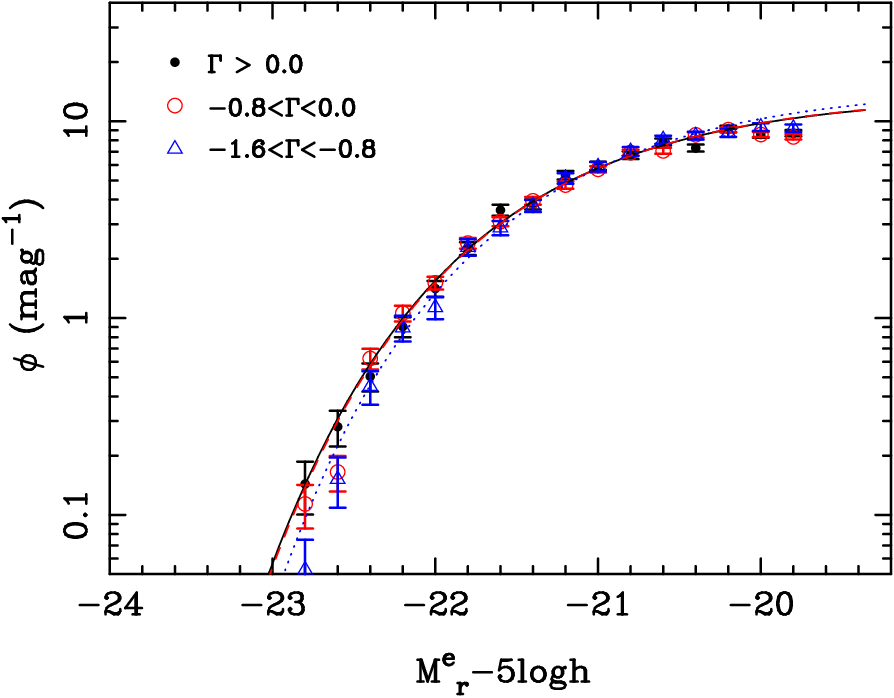}
\includegraphics[width = 72mm]{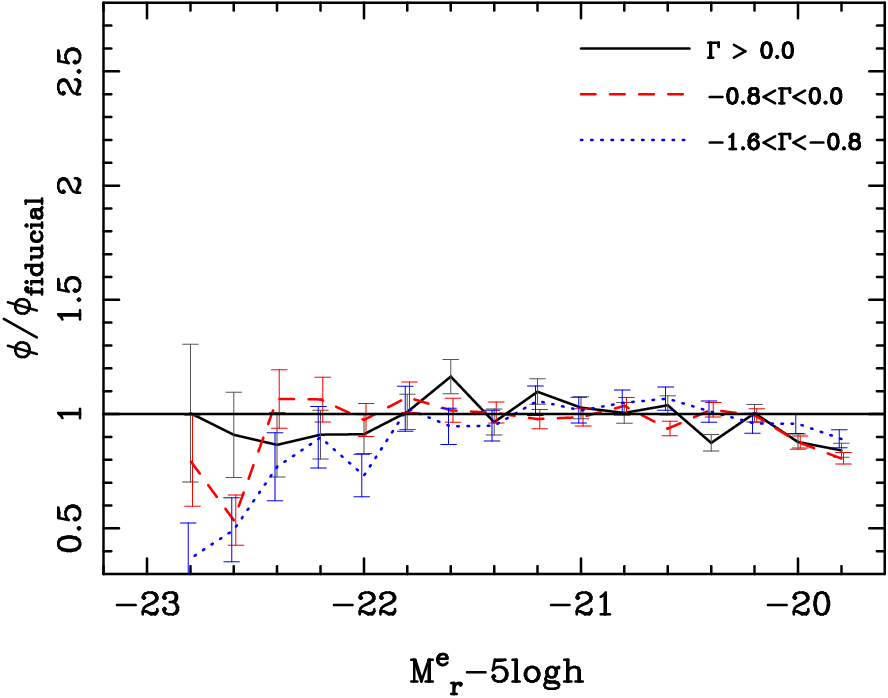}
\caption{Similar with Figure~\ref{relax}, but for galaxies in the
outer cluster region of $r_{500}$--$r_{200}$.}
\label{500to200}
\end{figure}

\subsection{Luminosity function of non-BCGs in clusters}

The sample of 2092 rich clusters of richness $R_{L\ast} \ge 50$ are
divided into three subsamples according to their dynamical states
quantified by relaxation parameter, $\Gamma$. The richness
distributions of these subsamples are very similar (see
Figure~\ref{relax_rich}), so that there is no selection effect on
richness in three subsamples.

We first calculate the composite luminosity function of non-BCG member
galaxies within the central region of $r_{500}=2/3r_{200}$
\citep{sks+03}, and fit them with the Schechter function (see the
upper panel of Figure~\ref{relax}). We only obtain the bright end of
galaxy luminosity function, which is insensitive to the faint end
slope $\alpha$. Hence, we fix $\alpha=-1.0$
\citep[e.g.,][]{pbr+05,lmg+06,dpb13} in the fitting, and compare
$M_{\ast}$ for clusters in different range of $\Gamma$. The derived
parameters, $\phi_{\ast}$ and $M_{\ast}$ are given in
Table~\ref{tab1}. We find that the luminosity functions at $M^e_r -
5\log h > -21.0$~mag agree with each other for different ranges of
$\Gamma$, but there is a significant excess for more unrelaxed (i.e.,
lower $\Gamma$) clusters at the bright end of $M^e_r -5\log h
<-21.0$~mag. Thus, more relaxed clusters have a fainter
$M_{\ast}$. The value of $M_{\ast}$ for relaxed clusters of
$\Gamma>0.0$ is 0.27 magnitude fainter than that for the unrelaxed
clusters of $-1.6<\Gamma<-0.8$.
To clearly show the excess of bright galaxies, we take the best-fit
Schechter function of galaxies in relaxed clusters of $\Gamma>0.0$ as
a fiducial line, and compare the ratios of luminosity functions to
this fiducial line (lower panel of Figure~\ref{relax}). Obviously, the
ratio of luminosity function of galaxies in unrelaxed clusters
significantly increases at $M^e_r-5\log<-21.0$~mag, which means that
there are more bright member galaxies in more unrelaxed clusters.

For a comparison, we also obtain the bright end of luminosity
functions of galaxies in the outer cluster region between $r_{500}$
and $r_{200}$ for clusters in the three relaxation parameter
ranges. As shown in Figure~\ref{500to200}, the luminosity functions of
these outer galaxies are very consistent for clusters with various
dynamical states, even at the bright end of $M^e_r -5\log h
<-21.0$~mag. This is inconsistent with the result of \citet{bgb+12}
who found a larger difference of galaxy population in the outer
cluster region at the bright end. We therefore can conclude that more
relaxed clusters have fewer bright member galaxies within $r_{500}$,
but in the outer cluster region ($>r_{500}$) the luminosity
distribution of member galaxies is nearly independent of cluster
dynamical state.

\begin{figure}
\centering
\includegraphics[width = 75mm]{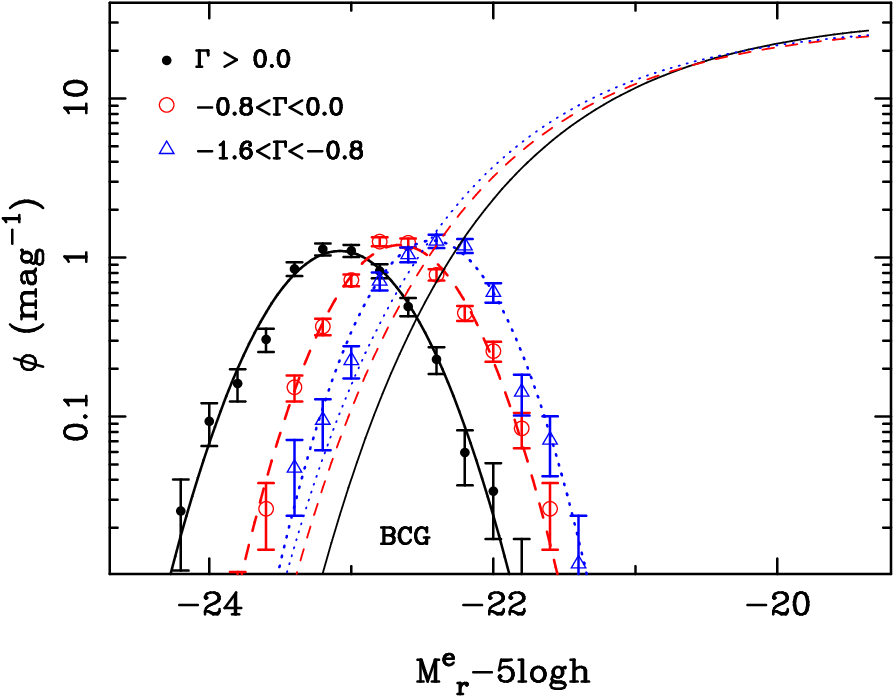}
\caption{Composite luminosity functions of BCGs and the best-fit
  Schechter functions of non-BCG member galaxies (thin lines) for
  clusters in the three ranges of relaxation parameters.}
\label{bcg}
\end{figure}

\subsection{Luminosity function of BCGs}
\label{bcglf}

The BCG in a galaxy cluster is the most massive galaxy near the center
of the cluster. The BCGs of many clusters have different statistical
properties from the non-BCG member galaxies
\citep{vbk+07,bhs+07,lwh+12,sym+14}. It has been suggested that the
BCGs were formed at redshift $z>2$, and then evolved passively
\citep[e.g.][]{ses+08,wad+08}. However, some simulation shows that
the BCGs were formed by accretion and merger of satellite galaxies
\citep[e.g.][]{db07}. The BCGs are very bright, and have a similar
absolute magnitude with a small dispersion of $\sim$0.3--0.45 mag
\citep{san88,abk98,lms04}. The composite luminosity function of the
BCGs can be described by a Gaussian function \citep{hmw+05}:
\begin{equation}
\phi_g(M)dM=\frac{\phi_0}{\sqrt{2\pi}\sigma_0}\exp\Big[{-\frac{(M-M_0)^2}{2\sigma^2_0}}\Big]dM,
\end{equation}
where $\phi_0$ is the normalization, $M_0$ and $\sigma_0$ are the mean
magnitude and the magnitude dispersion, respectively.

For clusters in the three ranges of $\Gamma$, we get three composite
BCG luminosity functions (see Figure~\ref{bcg}), and obtain the fitted
parameters in Table~\ref{tab1}. The dispersion of BCG absolute
magnitude is $\sim$0.36. In contrast to the non-BCG member galaxies,
we find that more relaxed clusters have a brighter BCG, e.g.
$M_0-5\log h = -23.08\pm0.04$ for the relaxed clusters of
$\Gamma>0.0$, compared to $M_0 -5\log h = -22.43\pm0.02$ for the very
unrelaxed clusters of $-1.6<\Gamma<-0.8$.

\section{Discussions and conclusions}

The total composite luminosity function of member galaxies 
in clusters should be the summation of $\phi_s$ and $\phi_g$, as 
\begin{equation}
\phi_{\rm tot}(M)dM=[\phi_s(M)+\phi_g(M)]dM.
\end{equation}
By using 2092 rich clusters, the largest sample of galaxy clusters
with quantified dynamical state, we find different dependence of
$\phi_s$ and $\phi_g$ for bright member galaxies on cluster dynamical
state. This is a clear evidence for the co-evolution of bright member
galaxies with cluster dynamical state. The mean absolute magnitude of
BCGs in clusters varies about 0.65~mag for different dynamical
states, while the characteristic magnitude $M_{\ast}$ of the non-BCG
member galaxies varies only about 0.27~mag. Note, however, that the
above results are obtained for the bright galaxies in the inner region
of clusters of $r<r_{500}$.  The luminosity function of bright member
galaxies in the outer region does not show dependence on cluster
dynamical state, which is consistent with the conclusion given by
\citet{dcd+03} and \citet{dpb13} who found the independence of galaxy
population on cluster dynamical state. Our conclusion is opposite to
that given by \citet{bgb+12} who showed the more significant
dependence of galaxy luminosity function in the outer cluster region
than that in the inner region.

How to explain the obvious difference of bright member galaxies in
clusters with different dynamical states? During relaxation process of
a cluster, many massive galaxies tend to sink to the center of a
cluster due to dynamical friction, and may be merged into the BCG
which produces a brighter BCG finally. This causes fewer bright
non-BCG member galaxies in the inner region of clusters. Observations
have showed that the BCGs in some clusters are experiencing major
merger \citep{mgh+08,lmd+09}. More relaxed clusters have a larger
magnitude gap between the first-rank and second-rank BCGs
\citep{rbp+07,skd+10,wh13}. Our results indicate that the evolution of
massive cluster galaxies deviates from a simple pure passive evolution
model, and somewhat support the scenario of hierarchical formation of
the BCGs \citep{db07}.

In summary, we study the dependence of the bright end of galaxy
luminosity function on cluster dynamical state by using the bright
member galaxies of a large sample of clusters. After a
redshift-evolution correction for the absolute magnitude of galaxies,
the composite luminosity function of non-BCG member galaxies can be
well fitted by the Schechter function. The absolute magnitude of BCGs
follows a Gaussian function with a dispersion of about
0.36~mag. Though in the outer cluster region ($>r_{500}$) the
luminosity function of bright member galaxies is independent of
cluster dynamical state, we find that in the cluster central region of
$r_{500}$, luminosity function of more relaxed clusters has a fainter
$M_{\ast}$. In these relaxed clusters, there are fewer bright member
galaxies of $M^e_r<-21.0+5\log h$ but have a brighter BCG. Our results
suggest the co-evolution of member galaxies with cluster dynamical
state and somewhat support the hierarchical formation scenario of the
BCGs.

\section*{Acknowledgments}

We thank the referee for valuable comments that helped to improve the
paper. The authors are supported by the National Natural Science
Foundation of China (11103032 and 11473034) and by the Strategic
Priority Research Program ``The Emergence of Cosmological Structures''
of the Chinese Academy of Sciences, Grant No. XDB09010200”.
Funding for SDSS-III has been provided by theAlfred P. Sloan
Foundation, the Participating Institutions, the National Science
Foundation, and theUSDepartment of Energy.  The SDSS-III Web site is
http://www.sdss3.org/.
SDSS-III is managed by the Astrophysical Research Consortium for the
Participating Institutions of the SDSS-III Collaboration including the
University of Arizona, the Brazilian Participation Group, Brookhaven
National Laboratory, University of Cambridge, University of Florida,
the French Participation Group, the German Participation Group, the
Instituto de Astrofisica de Canarias, the Michigan State/Notre
Dame/JINA Participation Group, Johns Hopkins University, Lawrence
Berkeley National Laboratory, Max Planck Institute for Astrophysics,
NewMexico State University, New York University, Ohio StateUniversity,
Pennsylvania State University,University of Portsmouth, Princeton
University, the Spanish Participation Group, University of Tokyo,
University of Utah,Vanderbilt University, University of Virginia,
University of Washington, and Yale University.

\label{lastpage}
\end{document}